\documentclass[12pt]{article}

\usepackage{graphicx}
\usepackage{amsmath}
\usepackage{amssymb}
\usepackage{bbm}
\usepackage{hyperref} 
\usepackage{natbib}
\usepackage{psfrag}
\usepackage{epsf}
\usepackage{url} 
\usepackage{enumerate}

\newcommand{\blind}{0}

\newtheorem{theorem}{Theorem}[section]

\newtheorem{proposition}[theorem]{Proposition}

\newcommand{\qed}{\nobreak \ifvmode \relax \else
      \ifdim\lastskip<1.5em \hskip-\lastskip
      \hskip1.5em plus0em minus0.5em \fi \nobreak
      \vrule height0.75em width0.5em depth0.25em\fi}

\def\bs{\boldsymbol}

\begin{document}

\def\spacingset#1{\renewcommand{\baselinestretch}%
{#1}\small\normalsize} \spacingset{1}

\if0\blind
{
  \title{\bf Testing Sparsity-Inducing Penalties}
  \author{Maryclare Griffin\thanks{
    The authors gratefully acknowledge support from NSF grants DGE-1256082 and DMS-1505136.}\hspace{.2cm}\\
    Department of Statistics, University of Washington\\
    and \\
    Peter D. Hoff \\
    Department of Statistical Science, Duke University}
  \maketitle
} \fi

\if1\blind
{
  \bigskip
  \bigskip
  \bigskip
  \begin{center}
    {\LARGE\bf Testing Sparsity-Inducing Penalties}
\end{center}
  \medskip
} \fi


\bigskip
\begin{abstract} 
Many penalized maximum likelihood estimators correspond to posterior mode estimators under specific prior distributions.
Appropriateness of a particular class of penalty functions can therefore be interpreted as the appropriateness of a prior  for the parameters. For example, the appropriateness of a lasso penalty for regression coefficients depends on the extent to which the empirical distribution of the regression coefficients resembles a Laplace distribution.
We give a testing procedure of whether or not a Laplace prior  is appropriate and accordingly, whether or not using a lasso penalized estimate is appropriate. 
This testing procedure is designed to have power against exponential power priors which correspond to $\ell_q$ penalties.
Via simulations, we show that this testing procedure achieves the desired level and has enough power to detect violations of the Laplace assumption when the numbers of observations and unknown regression coefficients are large. 
We then introduce an adaptive procedure that chooses a more appropriate prior and corresponding penalty from the class of exponential power priors when the null hypothesis is rejected. 
We show that this can improve estimation of the regression coefficients both when they are drawn from an exponential power distribution and when they are drawn from a spike-and-slab distribution.  
\end{abstract}

\vfill

\newpage 
\spacingset{1.45} 

\section{Introduction}

Lasso estimators are ubiquitous in linear regression due to their desirable properties and computational feasibility, as they can be used to produce sparse estimates of regression coefficients without sacrificing convexity of the estimation problem \citep{Tibshirani2011}.
The lasso estimator solves $\text{min}_{\bs \beta} \left|\left|\bs y - \bs X \bs \beta \right|\right|^2_2 + \lambda \left|\left|\bs \beta\right|\right|_1$, where $\bs y$ is an $n\times 1$ vector of responses, $\bs X$ is an $n\times p$ design matrix and the value $\lambda > 0$ determines the relative importance of the penalty $\left|\left|\bs \beta\right|\right|_1$ compared to the model fit $\left|\left|\bs y - \bs X \bs \beta\right|\right|^2_2$ in estimating $\bs \beta$. 
It has long been recognized that the lasso estimator corresponds to the posterior mode estimator when $\bs y = \bs X \bs \beta + \bs z$ and elements of $\bs z$ and $\bs \beta$ are independent normal and Laplace random variables, respectively \citep{Tibshirani1996,Figueiredo2003}.
The Laplace prior interpretation is popular in part because sampling from the full posterior distribution using a Gibbs sampler is computationally feasible \citep{Park2008}.
This allows computation of alternative posterior summaries, e.g. the posterior mean, median and quantiles, which can be used to obtain point and interval estimates of $\bs \beta$.
Furthermore, it has long been known that the interpretation of a penalty as a prior distribution yields decision theoretic justifications for using the corresponding penalized estimator, posterior mean estimator or posterior median estimator of $\bs \beta$ \citep{Pratt1965, Tiao1973}.  
For such reasons the interpretation of a penalty as a prior distribution has proliferated in many fields, especially in genetics research \citep{Legarra2011, Niemi2015, Leday2017}.

However, many researchers have found that the lasso estimator may perform suboptimally compared to other penalized estimators if the true value of $\bs \beta$ is highly sparse or not sparse at all \citep{Fan2001, Leeb2008}.
Analogously, posterior summaries under a Laplace prior have been found to be suboptimal compared to posterior summaries under other priors, depending on the empirical distribution of true values of the elements of $\bs \beta$ \citep{Griffin2010,Carvalho2010,Castillo2015,Bhattacharya2015,VanderPas2017}.
As a result, although a lasso estimator or a Laplace prior may be a reasonable default choice in high-dimensional problems due to its blend of desirable properties and computational feasibility, it would be useful to have a data-driven means to assess its appropriateness, and choose a more appropriate prior or penalty if suggested by the data.

The development of data-driven approaches to assess the appropriateness of priors for regression coefficients has long been a feature of the mixed model literature, in which regression coefficients are often modeled as normal random variables but more complex priors might be more appropriate. For instance, \cite{Verbeke1996} and \cite{Zhang2008} introduced alternative distributions for random effects that generalize the normal distribution, and \cite{Claeskens2009} and \cite{Drikvandi2016} developed novel tests of the appropriateness of normal priors for $\bs \beta$ that can be used to determine whether or not a normal prior for $\bs \beta$ and accordingly, simpler computation, is appropriate.

We take a similar approach, and assume the more flexible class of exponential power prior distributions which includes the Laplace prior as a special case  \citep{Subbotin1923, Box1973}:
 \begin{align}\label{eq:mod}
&\bs y =  \bs X \bs \beta + \bs z\text{, \quad} \beta_1,\dots, \beta_p \stackrel{i.i.d.}{\sim} EP\left(\tau, q\right)\text{, \quad} \bs z \sim N\left(\bs 0,\sigma^2 \bs I_n\right),
\end{align}
where $\sigma^2$ is the variance of the error $\bs z$ and $EP$ is the exponential power distribution with unknown shape parameter $q > 0$, parameterized such that $\tau^2$ is the variance of the unknown regression coefficients.
The corresponding posterior mode estimator of $\bs \beta$ is an $\ell_q$ penalized estimate which solves $\text{min}_{\bs \beta} \left|\left|\bs y - \bs X \bs \beta \right|\right|^2_2 + \lambda \left|\left|\bs \beta\right|\right|^q_q$ for some $q > 0$, where $\left|\left|\bs \beta\right|\right|^q_q = \sum_{j = 1}^p \left|\beta_j\right|^q$ and $\lambda = \tau^{-q} (\frac{\Gamma\left(3/q\right)}{\Gamma\left(1/q\right)} )^{q/2}$.
This includes the ridge estimator given by $q = 2$, which has long been known to have desirable shrinkage properties \citep{Hoerl1970}.
The $\ell_q$ class includes the class of bridge estimators described by \cite{Frank1993} and accordingly, penalties that can outperform $\ell_1$ penalties when the true value of $\bs \beta$ is highly sparse, at the cost of losing convexity of the estimation problem \citep{Huang2008, Mazumder2011, Marjanovic2014}. 
Posterior simulation under an exponential power prior can also be more computationally demanding, however the corresponding posterior summaries may outperform those based on Laplace priors for highly sparse or non-sparse $\bs \beta$.

When considering an exponential prior or $\ell_q$ penalty for $\bs \beta$, the question of whether or not a Laplace prior is appropriate is equivalent to the question of whether or not $q = 1$ is appropriate.
Without treating the $\ell_q$ penalty as indicating a model for $\bs \beta$, we might evaluate the appropriateness of $q = 1$ using cross validation, generalized cross validation or unbiased risk estimate minimization to choose values of $\lambda$ and $q$ simultaneously. 
However, these procedures can be challenging to perform over a two dimensional grid. 
Alternatively, fully Bayesian inference could proceed by assuming priors for the error variance $\sigma^2$, $\lambda$ and $q$ at the expense of losing a possibly sparse and computationally tractable posterior mode estimator \citep{Polson2014}. Accordingly, we could evaluate the appropriateness of a Laplace prior by computing a Bayes factor. However, specifying reasonable priors for $\lambda$ and $q$ that yield a proper posterior distribution is difficult in practice \citep{Fabrizi2010, Salazar2012}.  
Another option might be to construct a likelihood ratio or Wald test of the null hypothesis $H:q = 1$ against the alternative hypothesis, $K:q\neq 1$ under the model given by \eqref{eq:mod}.
However, constructing a likelihood ratio test would require prohibitively computationally demanding maximum marginal likelihood estimation of $\tau^2$, $\sigma^2$ and $q$ under the alternative, as well as derivation of the distribution of the test statistic under the null. 
Maximum marginal likelihood estimation of $\tau^2$, $\sigma^2$ and $q$ is challenging for several reasons. The form of the marginal likelihood as a function of $q$ and $\bs \beta$ is not amenable to an EM algorithm. Furthermore, approximations to the marginal likelihood may be difficult to construct because of the exponential power prior density is not differentiable as a function of $\bs \beta$ when $q \leq 1$ and, even if available, approximations to the marginal likelihood may perform poorly when $n$ is large relative to $p$, which is when it is most beneficial to assume a Laplace prior \citep{Barber2016, Huri2016}.
At the same time, derivation of the distribution of the test statistic under the null is challenging due to marginal dependence of $\bs y$ induced by assuming a prior distribution for $\bs \beta$. 
Importantly, all of these approaches also share the disadvantage of requiring penalized estimation or posterior simulation for $q \neq 1$ be performed regardless of whether or not $q = 1$ is deemed appropriate. This negates the computational advantages offered by assuming $q = 1$, as it requires penalized estimation or posterior simulation for $q \neq 1$ even when a test fails to reject the null hypothesis that $q= 1$ is appropriate.

In this paper we consider the Laplace and exponential power prior interpretations of the lasso and $\ell_q$ penalties and propose fast and easy-to-implement procedures for testing the appropriateness of a Laplace prior ($q = 1$) that do not require penalized estimation or posterior simulation for $q\neq 1$ as well as procedures for estimating $q$ and accordingly $\bs \beta$ in the event of rejection.
In Section~\ref{sec:test} we describe our testing procedure, which rejects the Laplace prior if an estimate of the kurtosis of the elements of $\bs \beta$ exceeds a particular threshold. 
The threshold is chosen so that the test rejects with probability approximately equal to $\alpha$, on average across datasets and Laplace-distributed coefficient vectors $\bs \beta$. 
We evaluate the performance of the approximation and the power of the testing procedure numerically via simulation.
In Section~\ref{sec:adapt} we introduce moment-based empirical Bayes estimates of $q$ and the variances $\sigma^2$ and $\tau^2$ of the error and the regression coefficients. We also propose a two-stage adaptive procedure for estimating $\bs \beta$.
If the testing procedure accepts the null, the adaptive estimation procedure defaults to an estimate computed under a Laplace prior. 
Otherwise, we estimate $\bs \beta$ under an exponential power prior using an estimated value of $q$. 
Because it is well known that a Laplace prior can yield suboptimal estimates of $\bs \beta$, we also compare the adaptive estimation procedure to estimates based on a Dirichlet-Laplace prior, which has been shown to outperform estimates based on a Laplace prior in certain settings \citep{Bhattacharya2015}.
We show via simulation that the adaptive estimation procedure outperforms estimators based on Laplace and Dirichlet-Laplace priors when elements of $\bs \beta$ have an exponential power distribution with $q< 1$, performs similarly to estimators based on a Laplace or Dirichlet-Laplace priors when elements of $\bs \beta$ have a Laplace distribution and outperforms estimators 
In Section~\ref{sec:sparse}, we demonstrate that the adaptive procedure also improves estimation of sparse $\bs \beta$ when elements of $\bs \beta$ have a spike-and-slab distribution.
In Section~\ref{sec:app}, we apply the testing and estimation procedures to several datasets commonly used in the penalized regression literature. A discussion follows in Section~\ref{sec:disc}.

\section{Testing the Laplace Prior}\label{sec:test}

\begin{figure}[h]
\centering
\includegraphics{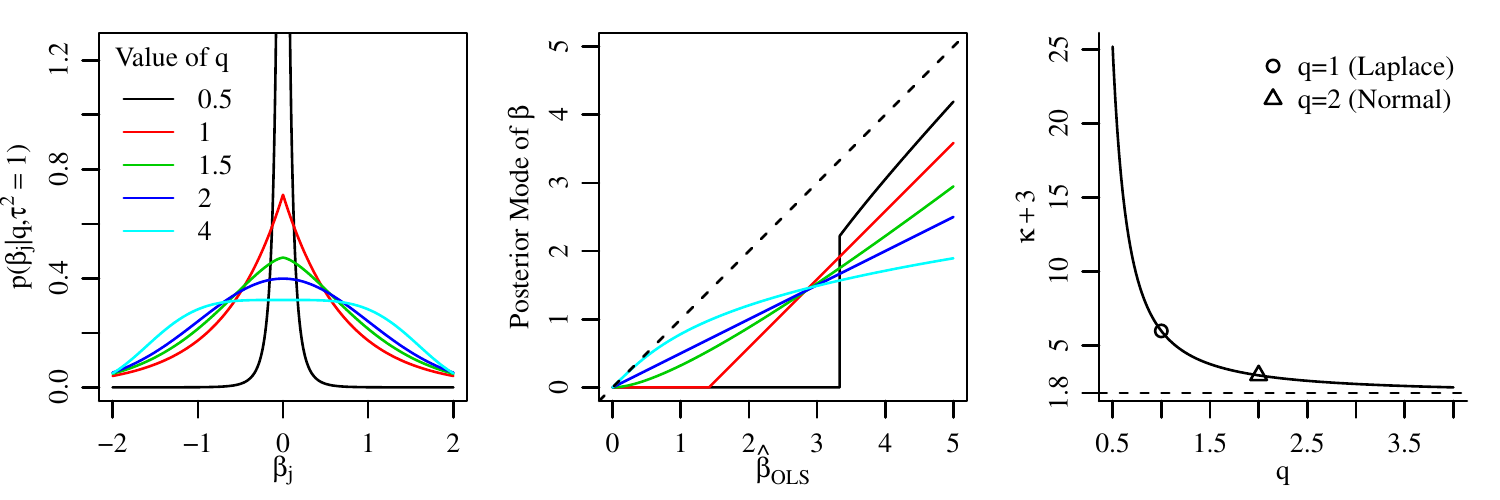}
\caption{The first panel shows exponential power densities for fixed variance $\tau^2 = 1$ and varying values of the shape parameter $q$. The second panel shows the mode thresholding function for $\sigma^2 = \tau^2 = 1$ and the values of $q$ considered in the first panel. The third panel shows the relationship between the kurtosis of the exponential power distribution and the shape parameter, $q$.}
\label{fig:powreg}
\end{figure}

Our approach to testing the appropriateness of the Laplace prior treats the Laplace prior as a special case of the larger class of exponential power distributions \citep{Subbotin1923, Box1973}. This class includes the normal and Laplace distributions.
The exponential power density is given by
\begin{align}\label{eq:epdens}
p\left(\beta_j | \tau^2, q\right) = \left(\frac{q}{2\tau}\right)\sqrt{\frac{\Gamma\left(3/q\right)}{\Gamma\left(1/q\right)^3}} \text{exp}\left\{- \left(\frac{\Gamma\left(3/q\right)}{\Gamma\left(1/q\right)} \right)^{q/2} \left|\frac{\beta_j}{\tau}\right|^q\right\},
\end{align}
where $q > 0$ is an unknown shape parameter and $\tau^2$ is the variance. The first panel of Figure~\ref{fig:powreg} plots exponential power densities for different values of $q$ with the variance fixed at $\tau^2 = 1$.
Because the exponential power distributions have simple distribution functions with easy to compute moments and can accommodate a wide range of tail behaviors, they quickly became popular as an alternative error distributions \citep{Subbotin1923,Diananda1949, Box1953, Box1973}.

When an exponential power prior is assumed for $\bs \beta$, we can understand how the choice of $q$ provides flexible penalization by examining the mode thresholding function. The mode thresholding function relates the OLS estimate for a simplified problem with a single standardized covariate to the posterior mode estimator of $\bs \beta$. Let $\bs x$ be a standardized $n\times 1$ covariate vector with $\left|\left|\bs x\right|\right|^2_2 = 1$, let $\beta$ be a scalar and let $\hat{\beta}_{ols} =\bs x^\top\bs y$. The mode thresholding function is given by:
\begin{align*}
\text{arg min}_{\beta} \frac{1}{2\sigma^2}\left(\hat{\beta}_{ols} - \beta\right)^2 + \left(\frac{\Gamma\left(3/q\right)}{\Gamma\left(1/q\right)} \right)^{q/2} \left|\frac{\beta}{\tau}\right|^q.
\end{align*}
This function is not generally available in closed form but can be computed numerically, even when $q < 1$ and the mode thresholding problem is non-convex \citep{Marjanovic2014}.
The second panel of Figure~\ref{fig:powreg} shows the mode thresholding function for $\sigma^2 = \tau^2 = 1$.

Within the class of exponential power priors, the relationship between the shape parameter $q$ and kurtosis $\mathbb{E}\left[\beta_j^4\right]/\mathbb{E}\left[\beta_j^2\right]^2$ is one-to-one and given by:
\begin{align}\label{eq:kurt}
	\kappa + 3= \Gamma\left(5/q\right)\Gamma\left(1/q\right)/\Gamma\left(3/q\right)^2,
\end{align} 
where $\kappa$ refers to the excess kurtosis relative to a normal distribution.
We plot kurtosis as a function of $q$ in the second panel of Figure~\ref{fig:powreg}.
Accordingly, if $\bs \beta$ were observed we could naively construct a test statistic based on the empirical kurtosis of the elements of $\bs \beta$. We define the test statistic $\psi(\bs \beta) = m_4\left(\bs \beta\right)/m_2\left(\bs \beta\right)^2$, where $m_2\left(\bs \beta\right) = \frac{1}{p}\sum_{j = 1}^p \beta^2_p$ and $m_4\left(\bs \beta\right)= \frac{1}{p}\sum_{j = 1}^p \beta^4_p$ are the second and fourth empirical moments of $\bs \beta$. The test statistic $\psi(\bs \beta)$ is the empirical kurtosis of the elements of the vector $\bs \beta$.
An exact level-$\alpha$ test of $H$ could be performed by comparing the test statistic to the $\alpha/2$ and $1 - \alpha/2$ quantiles of the distribution of $\psi(\bs \beta)$ under the null. Because the distribution of $\psi(\bs \beta)$ under an exponential power prior depends only on $q$ and not $\tau^2$, we can obtain Monte Carlo estimates of the $\alpha/2$ and $1 - \alpha/2$ quantiles $\psi_{\alpha/2}$ and $\psi_{1 - \alpha/2}$ by simulating entries of $\bs \beta^*$ from any Laplace distribution and and computing $\psi\left(\bs \beta^*\right)$.
As this test is available \emph{only} when $\bs \beta$ is observed, we refer to this as the oracle test.

In practice, $\bs \beta$ is not observed.  However when $n > p$ and $\bs X^\top\bs X$ is full rank, the OLS estimator $\hat{\bs \beta}_{ols} = \left(\bs X^\top \bs X\right)^{-1}\bs X^\top \bs y$ is available. As a surrogate for $\psi(\bs \beta)$, we can use $\psi(\hat{\bs \beta}_{ols})$ as a test statistic. If $n >> p$, then $\hat{\bs \beta}_{ols} \approx \bs \beta$ conditional on $\bs \beta$. It follows that $\psi\left(\hat{\bs \beta}_{ols}\right) \stackrel{d}{\approx} \psi(\bs \beta)$ when treating $\bs \beta$ as random.

\begin{proposition}\label{prop:olstest}
Under normality of the errors $\bs z$ as assumed in \eqref{eq:mod}, 
\begin{align}\label{eq:convols}
\mathbb{E}\left[(\psi(\hat{\bs \beta}_{ols}) - \psi(\bs \beta))^2 | \bs \beta\right] \leq 16\sigma^2\left(\frac{m_6(\bs \beta)}{m_2(\bs \beta)^4}\right)\text{tr}((\bs X^\top\bs X)^{-1})/p + o\left(\sigma^2\text{tr}((\bs X^\top\bs X)^{-1})/p \right)
\end{align}
where $m_6\left(\bs \beta\right) = \frac{1}{p}\sum_{j = 1}^p \beta_j^6$. 
\end{proposition} 
Details are provided in the appendix.
Accordingly, when $\text{tr}((\bs X^\top\bs X)^{-1})/p$ is small an approximate level-$\alpha$ test of $H$ is obtained by rejecting $H$ when $\psi\left(\hat{\bs \beta}_{ols}\right) \notin \left(\psi_{\alpha/2}, \psi_{1 - \alpha/2}\right)$.

Although the behavior of the OLS estimator is well understood, we introduce some additional notation to help explain when $\text{tr}( (\bs X^\top\bs X))^{-1}$ is likely to be small for large $n$.
Let $\bs V$ be a diagonal matrix with diagonal elements $\sqrt{\text{diag}\left(\bs X^\top\bs X\right)}$ and $\bs C$ be the ``correlation'' matrix corresponding to $\bs X^\top\bs X$, such that $\bs X^\top\bs X = \bs V \bs C \bs V$. Let $\eta_j$ refer to eigenvalues of $\bs C$. The eigenvalues $\eta_j$ indicate the overall collinearity of $\bs X$. When columns of $\bs X$ are orthogonal, $\eta_1 = \dots = \eta_p = 1$, whereas when $\bs X$ is highly collinear the smallest values of $\eta_j$ may be very close or exactly equal to $0$. 
Applying Theorem 3.4 of \cite{Styan1973} we can write 
\begin{align*}
\text{tr}((\bs X^\top\bs X)^{-1})/p \leq  \text{max}_j \left(\frac{1}{\left|\left|\bs x_j\right|\right|^2_2}\right) \text{max}_j \left(\frac{1}{\eta_j^2}\right).
\end{align*}
We can see that as long as $\left|\left|\bs x_j\right|\right|^2_2$ are large, which will tend to be the case when $n$ is large, and eigenvalues of $\eta_j$ are not very small, i.e.\ $\bs X$ is not too highly collinear, $\text{tr}((\bs X^\top\bs X)^{-1})/p$ will be small enough to justify using $\psi(\hat{\bs \beta}_{ols})$ as a surrogate for $\psi(\bs \beta)$. 

%

%


However, penalized regression is often considered when $n < p$ or $\bs X$ is highly collinear.
When $n < p$, the OLS estimator is not unique and so neither is $\psi(\hat{\bs \beta}_{ols})$. When $n \geq p$ but $\bs X$ is highly collinear, i.e.\ some columns of $\bs X$ are strongly correlated with others and $\text{tr}((\bs X^\top\bs X)^{-1})/p$ may not be small even for large values of $n$. When columns of $\bs X$ have been centered and standardized to have norm $n$ according to standard practice this is easy to see. The quantity $\text{tr}((\bs X^\top\bs X)^{-1})/p=\frac{1}{np}\sum_{j = 1}^p \frac{1}{\eta_j}$ will ``blow up'' if any values of $\eta_j$ are very close to or exactly equal to zero and quantiles of $\psi(\bs \beta)$ will poorly approximate quantiles of $\psi(\hat{\bs \beta}_{ols})$. 

Fortunately, we can construct a modified test using a ridge estimate of $\bs \beta$, $\hat{\bs \beta}_{\delta} = \bs V^{-1}(\bs C+ \delta^2 \bs I_p)^{-1}\bs V^{-1} \bs X^\top\bs y$, where $\delta \geq 0$ is a nonnegative constant. 
Ridge estimators reduce variance at the cost of yielding a biased estimate of $\bs \beta$, $\mathbb{E}[\hat{\bs \beta}_{\delta} | \bs \beta]= \bs V^{-1}\left(\bs C+ \delta^2 \bs I_p\right)^{-1}\bs C \bs V \bs \beta$. 
However letting $\bs \beta_\delta = \mathbb{E}[\hat{\bs \beta}_{\delta} | \bs \beta]$, the distribution of $\psi\left(\bs \beta_\delta\right)$ under an exponential power prior still only depends on $q$ and not $\tau^2$. Accordingly, if $\bs \beta_\delta$ were observed we could perform an exact level-$\alpha$ test of $H$ by comparing $\psi\left(\bs \beta_\delta\right)$ to Monte Carlo estimates of the $\alpha/2$ and $1 - \alpha/2$ quantiles $\psi_{\delta,\alpha/2}$ and $\psi_{\delta,1 - \alpha/2}$ obtained by simulating $\bs \beta^*$ from any Laplace distribution and computing $\psi(\bs \beta_\delta^*= \bs V^{-1}\left(\bs C+ \delta^2 \bs I_p\right)^{-1}\bs C \bs V \bs \beta^*)$. In practice, we can use $\psi(\hat{\bs \beta}_{\delta})$ as a surrogate for $\psi\left(\bs \beta_\delta\right)$ to obtain an approximate level-$\alpha$ test.

\begin{proposition}\label{prop:ridgetest}
Let $\bs \Sigma_{\delta} = \bs V^{-1} \left(\bs C+ \delta^2 \bs I_p\right)^{-1}\bs C \left(\bs C+ \delta^2 \bs I_p\right)^{-1}\bs V^{-1}$.
Under normality of the errors $\bs z$ as assumed in the model given by \eqref{eq:mod},
\begin{align}\label{eq:convridge}
\mathbb{E}\left[(\psi(\hat{\bs \beta}_{\delta}) - \psi(\bs \beta_{\delta}))^2| \bs \beta\right] \leq 
16\sigma^2\left(\frac{m_6\left(\bs \beta_\delta\right)}{m_2(\bs \beta_\delta)^4}\right)\text{tr}\left(\bs \Sigma_\delta\right)/p + o\left(\sigma^2 \text{tr}\left(\bs \Sigma_\delta\right)/p \right),
\end{align}
where $m_k\left(\bs \beta_\delta\right) = \frac{1}{p}\sum_{j = 1}^p \beta_j^k$.
\end{proposition}
Details are provided in the appendix.
It follows that when $\text{tr}\left(\bs \Sigma_\delta\right)/p$ is small, an approximate level-$\alpha$ test of $H$ is obtained by rejecting $H$ when $\psi(\hat{\bs \beta}_{\delta}) \notin\left( \psi_{\delta,\alpha/2}, \psi_{\delta,1 - \alpha/2}\right)$. 

As the performance of this test depends on $\text{tr}\left(\bs \Sigma_{\delta}\right)/p$, it depends not only on $\bs X$ but also on $\delta^2$.
Again applying Theorem 3.4 of \cite{Styan1973} we can write 
\begin{align*}
\text{tr}\left(\bs \Sigma_{\delta}\right)/p \leq \text{max}_j \left(\frac{1}{\left|\left|\bs x_j\right|\right|^2_2}\right)\text{max}_j \left(\frac{\eta_j}{\left(\eta_j + \delta^2\right)^2}\right).
\end{align*}
The first term depends only on the design matrix, $\bs X$. As long as $\left|\left|\bs x_j\right|\right|^2_2$ are large, which again is likely to be the case for large $n$, the first term will be small.
The second term depends on $\delta^2$ through the the eigenvalue ratios $\frac{\eta_j}{\left(\eta_j + \delta^2\right)^2}$.
Heuristically, the eigenvalue ratios are decreasing in $\delta^2$ and setting $\delta^2$ to be very large would ensure that $\text{tr}\left(\bs \Sigma_{\delta}\right)/p$ is very close to $0$ and that $\psi(\hat{\bs \beta}_\delta)$ performs well as a surrogate for $\psi(\bs \beta_\delta)$. However, increasing $\delta^2$ also reduces the power of the test as it forces the ridge estimate closer to the zero vector. 
To ensure that the eigenvalue ratios do not ``blow up'' while retaining as much power as possible we recommend setting $\delta^2 = \left(1-\text{min}_j \eta_j\right)_+$, where $\eta_1, \dots, \eta_p$ are the eigenvalues of $\bs C$.
When the columns of $\bs X$ are standardized to have norm $n$, $\text{tr}\left(\bs \Sigma_{\delta}\right)/p=\frac{1}{np} \sum_{i = 1}^p \frac{\eta_j}{\left(\delta^2 + \eta_j\right)^2}$. Accordingly with $\delta^2$ set to $\delta^2 = \left(1-\text{min}_j \eta_j\right)_+$, we can at least ensure that $\text{tr}\left(\bs \Sigma_{\delta}\right)/p \leq \frac{1}{n}$.

The tests based on $\psi(\hat{\bs \beta}_{ols})$ and $\psi(\hat{\bs \beta}_{\delta})$ have several good features. First, the approximate distributions of the test statistics $\psi(\hat{\bs \beta}_{ols})$ and $\psi(\hat{\bs \beta}_{\delta})$ do not depend on the values of the unknown parameters $\tau^2$ or $\sigma^2$, and so their approximate null distributions may be simulated easily.
Second, both test statistics are easy and quick-to-compute even for very high dimensional data. 
Third, both test statistics are invariant to rescaling of $\bs y$ or $\bs X$ by a constant.

\begin{figure}[h]
\centering
\includegraphics{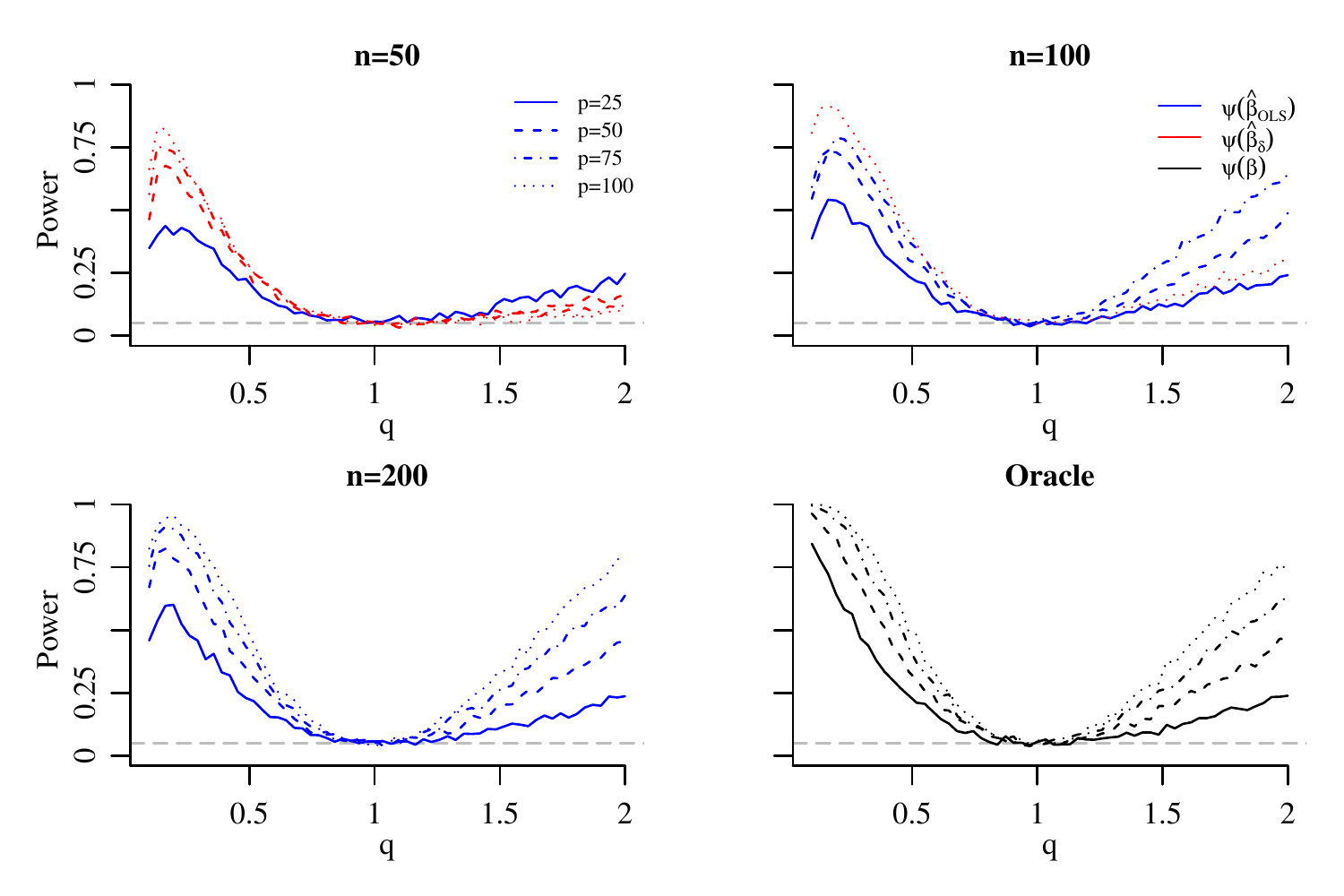}
\caption{Power and level of level-$0.05$ tests for data simulated from model \eqref{eq:mod} with exponential power distributed $\bs \beta$ and $\sigma^2 = \tau^2 = 1$. A horizontal dashed gray line is given at $0.05$.}
\label{fig:power_nondiag_pow}
\end{figure}

We examine the performance of the tests with a simulation study.
We simulate parameters and data according to the model \eqref{eq:mod}.
When simulating data and parameters, we set $\sigma^2 = \tau^2 = 1$ and consider  $p \in \left\{25, 50, 75, 100\right\}$, $n \in \left\{50, 100, 200\right\}$ and $q \in \left\{0.1,\dots,2\right\}$.  Because the OLS and ridge test statistics are invariant to rescaling of $\bs y$ by a constant, the simulation results depend only on $\tau^2/\sigma^2$, and in this case reflect the performance of the tests when $\tau^2 = \sigma^2$. For each combination of $p$, $n$ and $q$, we simulate $1,000$ values of $\bs X$ and $\bs \beta$, drawing entries of $\bs X$ independently from the standard normal distribution. 
When $n > p$, we use the OLS test statistic $\psi(\hat{\bs \beta}_{ols})$. When $n \leq p$, we use the ridge test statistic $\psi(\hat{\bs \beta}_{\delta})$.
Figure~\ref{fig:power_nondiag_pow} shows the power of the level-$0.05$ tests, i.e.\ the proportion of simulated datasets for which we reject $H$ at level-$0.05$ as a function of $q$ and $n$.
When $q = 1$, this gives the level of the test.
The last panel shows the power of the oracle test based on $\psi(\bs \beta)$.

The simulation results shown in Figure~\ref{fig:power_nondiag_pow} indicate that the tests will perform well relative to the oracle test for this range of values of $n$ and $p$.
The power of the test is increasing in $p$, as this in a sense represents our sample size for evaluating the distribution of $\bs \beta$.
The power of the test is also increasing in $q$ moves away from $q=1$, i.e. as the empirical distribution of the elements of $\bs \beta$ becomes less similar to a Laplace distribution.
As we might expect given that the ridge estimator corresponds to an estimator of $\bs \beta$ under a normal prior with variance $1/\delta^2$, using the modified ridge-based test results in a reduction of power especially against alternatives with $q > 1$. 
Interestingly, we see that the power of the test is not symmetric with respect to how far the true value of $q$ is from $1$. 
This is due to the fact that kurtosis is changing more slowly as a function of $q$ as $q$ increases, as can be seen in Figure~\ref{fig:powreg}.
We also observe a dip in power for very small values of $q$ when $\psi(\hat{\bs \beta}_{ols})$ or $\psi(\hat{\bs \beta}_{\delta})$ are used. This can be explained by examining the bias of $m_2(\hat{\bs \beta}_{ols})$ which appears in the denominator of $\psi(\hat{\bs \beta}_{ols})$
\begin{align*}
\mathbb{E}\left[m_2(\hat{\bs \beta}_{ols})\right] = m_2(\bs \beta) + \sigma^2 \text{tr}((\bs X^\top\bs X)^{-1})/p.
\end{align*}
When $q$ is very small and $p$ is less than or equal to $100$, most elements of $\bs \beta$ will be very close to zero with high probability.
For instance, when $q = 0.1$ and $\tau^2 = 1$,  $\text{Pr}(|\beta_j|\geq 0.1) \approx 0.08$. When most elements of $\bs \beta$ are very close to $0$, $m_2\left(\bs \beta\right)$ will be small and $m_2(\hat{\bs \beta}_{ols})$ will be dominated by the error incurred by estimating $\bs \beta$. The behavior of $\psi(\hat{\bs \beta}_{\delta})$ at small values of $q$ can be explained analogously.

\section{Adaptive Estimation of $\bs \beta$}\label{sec:adapt}

Rejection of the null hypothesis implies that the empirical distribution of the unobserved entries of $\bs \beta$ does not resemble a Laplace distribution. This suggests a two-stage adaptive procedure for estimating $\bs \beta$ that first tests the appropriateness of the Laplace prior and estimates $\bs \beta$ under a Laplace prior if the test accepts and estimates $\bs \beta$ under an exponential power prior otherwise. 
This procedure requires estimates of $\tau^2$ and $\sigma^2$ if the test accepts and $\tau^2$, $\sigma^2$ and $q$ if the test rejects, as well as procedures for computing a posterior mode estimator or simulating from the posterior distribution of $\bs \beta$ under an exponential power prior. We do not specify which posterior summary should be used to estimate $\bs \beta$ in general. 
It is well known that  different posterior summaries minimize different loss functions \citep{Hans2009}, and we view the choice of posterior summary as problem-specific.

We consider empirical Bayes (EB) estimation of $q$, $\tau^2$ and $\sigma^2$.
Estimating these parameters by maximizing the marginal likelihood of the data $\int p\left(\bs y | \bs X, \bs \beta,\sigma^2 \right)p\left(\bs \beta | \tau^2,q\right)d\bs \beta$ is difficult because the integral is not available in closed form for arbitrary values of $q$. 
The problem is not amenable to a Gibbs-within-EM algorithm for maximizing over $\sigma^2$, $\tau^2$ and $q$ jointly and  Gibbs-within-EM algorithms to obtain maximum marginal likelihood estimates of $\tau^2$ and $\sigma^2$ for fixed $q$ are computationally intensive and tend to be slow to converge \citep{Roy2016}. 
As a result, we consider easy and quick to compute moment-based EB estimators of $\sigma^2$, $\tau^2$ and $q$.
As moment estimators, they are more robust to misspecification of the prior and residual distributions than likelihood-based alternatives. 
Conveniently, the estimators for $\tau^2$ and $\sigma^2$ do not depend on $q$. This yields simple and interpretable comparisons of estimates of $\bs \beta$ computed under Laplace versus exponential power priors.

\subsection{Estimation of $q$}

The test statistics $\psi(\hat{\bs \beta}_{ols})$ and $\psi(\hat{\bs \beta}_{\delta})$ can be used to estimate $q$.
In the previous section, we demonstrated that an approximate test of $H$ could be obtained by using $\psi(\hat{\bs \beta}_{ols})$ as a surrogate for $\psi(\bs \beta)$ when $\text{tr}((\bs X^\top\bs X)^{-1})/p$ is small. Recall that the quantity $\psi(\bs \beta)$ is the empirical kurtosis of $\bs \beta$ and is defined as a function of the second and fourth empirical moments of $\bs \beta$, $m_2(\bs \beta)$ and $m_4(\bs \beta)$ . As $m_2(\bs \beta)\stackrel{p}{\rightarrow}\mathbb{E}[\beta^2_j]$ and $m_4(\bs \beta)\stackrel{p}{\rightarrow}\mathbb{E}[\beta^4_j]$ as $p\rightarrow \infty$, it follows from the continuous mapping theorem that $\psi(\bs \beta)\stackrel{p}{\rightarrow} \kappa + 3$ as $p\rightarrow \infty$, where $\kappa + 3$ is the  kurtosis of the distribution of elements of $\bs \beta$. 
Accordingly, we can use $\psi(\hat{\bs \beta}_{ols})$ directly as an estimator of the kurtosis $\kappa + 3$ when $\text{tr}((\bs X^\top\bs X)^{-1})$ is small \emph{and} $p$ is large.

When the ridge-based test statistic $\psi(\hat{\bs \beta}_{\delta})$ is used, estimation of $\kappa$ is less straightforward. Even if $m_2(\bs \beta_\delta) \stackrel{p}{\rightarrow} \mathbb{E}[m_2(\bs \beta_\delta)]$ and $m_4(\bs \beta_\delta) \stackrel{p}{\rightarrow} \mathbb{E}[m_4(\bs \beta_\delta)]$ as $p\rightarrow \infty$, the continuous mapping theorem implies  $\psi(\bs \beta_\delta) \stackrel{p}{\rightarrow} \left(\gamma \left(\kappa + 3\right) + \omega \right)/\alpha^2$
as $p \rightarrow \infty$,
where  $\alpha = \text{tr}(\bs X^\top\bs X \bs D^2\bs X^\top\bs X)/p$, $\gamma = \sum_{j = 1}^p\sum_{k = 1}^p (\bs D \bs X^\top\bs X)_{jk}^4/p$, $\omega= 3((\sum_{j = 1}^p (\bs X^\top\bs X \bs D^2 \bs X^\top\bs X)^2_{jj})/p - \gamma )$ and $\bs D = \bs V^{-1}\left(\bs C + \delta^2 \bs I_p \right)^{-1}\bs V^{-1}$. 
This suggests the follow bias correction
\begin{align}
\widehat{\kappa + 3} = \left(\frac{\alpha^2}{\gamma}\right)\left(\psi(\hat{\bs \beta}_{\delta}) - \frac{\omega}{\alpha^2} \right). 
\end{align}
Additional details are provided in the appendix.

Given an estimate of $\kappa + 3$, we estimate $q$ from \eqref{eq:kurt}, $\kappa + 3= \Gamma\left(5/q\right)\Gamma\left(1/q\right)/\Gamma\left(3/q\right)^2$ using Newton's method.

\subsection{Estimation of $\sigma^2$ and $\tau^2$}

Under the model given by \eqref{eq:mod}, the marginal mean and variance of the data $\bs y$ are given by $\mathbb{E}\left[\bs y\right] = 0$ and $\mathbb{V}\left[\bs y \right] = \bs X \bs X^\top \tau^2 + \sigma^2 \bs I_n$. 
We can estimate $\tau^2$ and $\sigma^2$ by solving:
\begin{align}\label{eq:varmod}
\text{min}_{\tau^2, \sigma^2} \text{log}\left(\left|\bs X\bs X^\top \tau^2 + \bs I_n \sigma^2\right|\right) + \text{tr}\left(\bs y \bs y^\top \left(\bs X \bs X^\top \tau^2 + \bs I_n \sigma^2 \right)^{-1} \right).
\end{align}
Intuitively, this provides moment-based estimates of $\tau^2$ and $\sigma^2$ by minimizing a loss function relating the empirical variance $\bs y \bs y^\top$ to the variance $\bs X \bs X^\top \tau^2 + \sigma^2 \bs I_n$ under the model \eqref{eq:mod}, while requiring positive definiteness of $\bs X \bs X^\top \tau^2 + \sigma^2 \bs I_n$.
\cite{Hoff2017} demonstrate that these estimates will be consistent for $\tau^2$ and $\sigma^2$ as $n$ and $p\rightarrow \infty$ even if the distribution of $\bs \beta$ is not normal. 
Solving \eqref{eq:varmod} has been treated thoroughly in the random effects literature \citep{Demidenko2013}. 
We caution that when $n < p$, the solution to \eqref{eq:varmod} can lie on the boundary of the parameter space at $\sigma^2 = 0$.

\subsection{Estimation of $\bs \beta$ Given $\tau^2$, $\sigma^2$ and $q$}\label{subsec:est_beta}

Given $\tau^2$, $\sigma^2$ and $q$, we can compute the posterior mode estimator of $\bs \beta$ using a coordinate descent algorithm that utilizes the mode thresholding function depicted in Figure~\ref{fig:powreg}. \cite{Fu1998} provided coordinate descent algorithms for $q \geq 1$ and \cite{Marjanovic2014} gave a coordinate descent algorithm for $q < 1$ that is guaranteed to converge to a local minimum under certain conditions on $\bs X$.
Details of the coordinate descent algorithm are given in the appendix.
We note that when $q < 1$, the posterior mode optimization problem is not convex and the mode may not be unique.
   

Alternative posterior summaries, e.g. the posterior mean or median of $\bs \beta$ under the model given by \eqref{eq:mod} can be approximated using a Gibbs sampler that simulates from the posterior distribution of $\bs \beta$.
For any value of $q > 0$ there is a uniform scale mixture representation of the exponential power distribution \citep{Walker1999}. If $\beta_j$ has an exponential power distribution, we can write $\beta_j | \gamma_j \sim \text{uniform}\left(-\Delta_j, \Delta_j\right)$, where $\Delta_j = \gamma_j^{1/q}\sqrt{\left(\frac{\Gamma\left(1/q\right)}{\Gamma\left(3/q\right)}\right)\left(\frac{\tau^2}{2}\right)}$ and $\gamma_j \sim \text{gamma}\left(\text{shape}=1 + 1/q,\text{rate}=2^{-q/2} \right)$.
To our knowledge, this representation has not been used to construct a Gibbs sampler when an exponential power prior is assumed for regression coefficients corresponding to an arbitrary design matrix $\bs X$. 
Using this representation, the full conditional distribution for $\bs \beta$ given $\bs \gamma$ is a truncated multivariate normal distribution and the full conditional distributions for elements of $\bs \gamma$ given $\bs \beta$ are independent translated exponential distributions. Full conditional distributions are given in the appendix.

\subsection{Simulation Results}\label{subsec:adaptive_simulations}

We assess the performance of the adaptive procedure via simulation.
We simulate data from \eqref{eq:mod} with $\tau^2 = \sigma^2 = 1$, $p = 100$, $n \in \left\{100, 200\right\}$ and $q\in \left\{0.25, 1, 4\right\}$ and entries of $\bs X$ drawn from a standard normal distribution. For each pair of values of $n$ and $q$, we simulate $100$ values of $\bs y$ from \eqref{eq:mod}. When $n > p$, we use $100$ different design matrices, $\bs X$, whereas when $n \leq p$ we fix the design matrix $\bs X$ so that some matrix calculations involving $\bs X$ can be precomputed. 
As noted previously, when $n \leq p$ the solution to the variance component estimation problem \eqref{eq:varmod} can lie on the boundary of the parameter space at $\sigma^2 = 0$. 
For the purposes of this simulation study, we require simulated datasets yield $\hat{\sigma}^2 \neq 0$.

We examine the performance of the adaptive procedure posterior mean estimators, which are known to minimize posterior squared error loss and accordingly allow for straightforward performance comparisons when entries of $\bs \beta$ are continuous \citep{Tiao1973}.
We approximate posterior mean estimators from  $10,500$ simulations from the posterior distribution using the Gibbs sampler described in Section~\ref{subsec:est_beta},  discarding the first $500$ iterations as burn-in. 
In general, the sampler mixes better with larger $q$ and $n$. The smallest effective sample sizes for $n = 100$ and $n = 200$ are $53$ and $222$, respectively.
Histograms of estimates of $\sigma^2$, $\tau^2$ and $q$ are given in the appendix.

\begin{figure}
\includegraphics{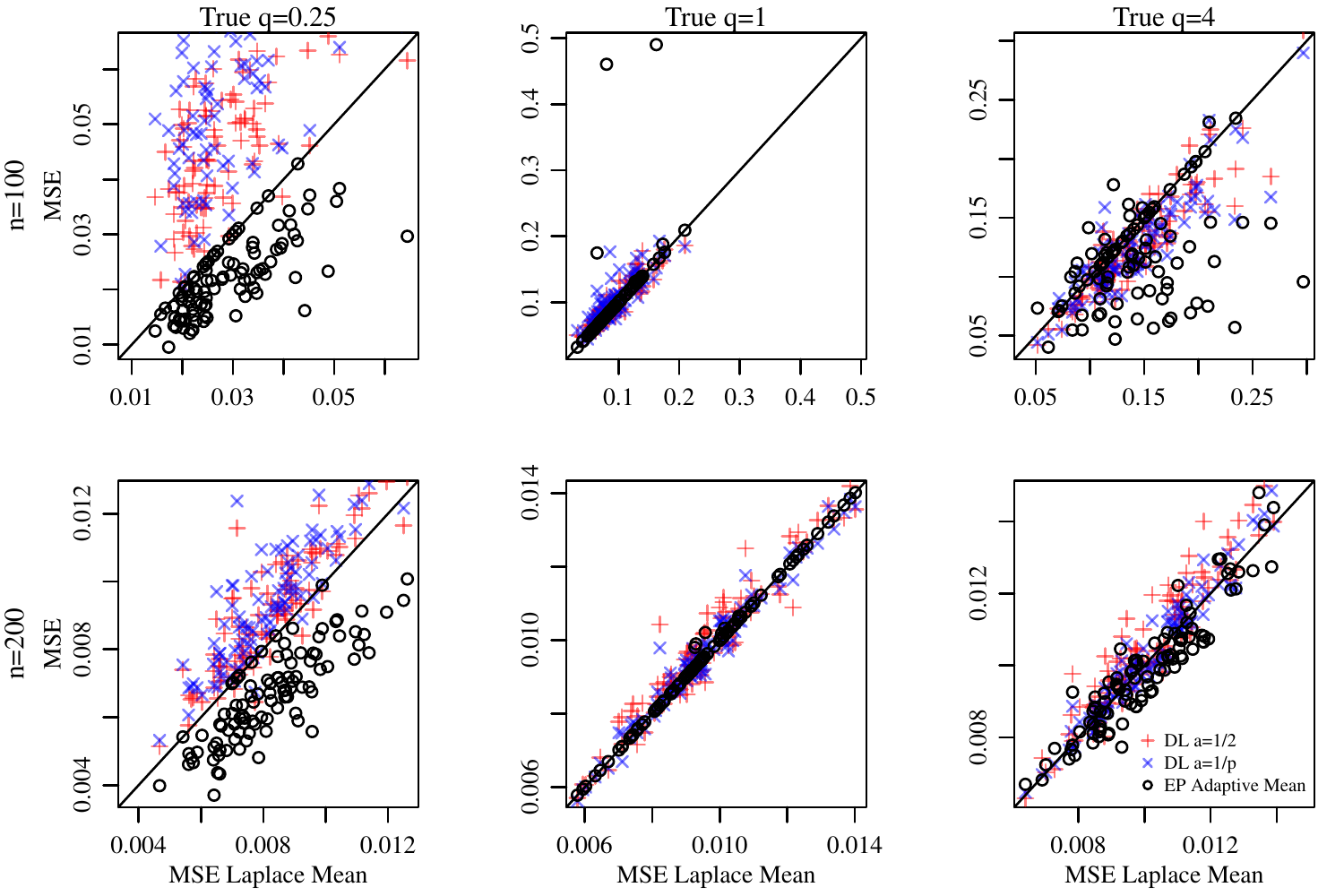}
\caption{Adaptive estimation procedure posterior mean estimator and Dirichlet-Laplace (DL) posterior mean estimator versus Laplace prior posterior mean estimator performance, as measured by mean squared error (MSE), for data simulated from model \eqref{eq:mod} with exponential power distributed $\bs \beta$ and $\sigma^2 = \tau^2 = 1$.}
\label{fig:compests_mse}
\end{figure}

For $n = 100$, we reject the null hypothesis that $q = 1$ at level $\alpha = 0.05$ in 100\%, 1\% and 62\% of the simulations when $q = 0.25$, $q = 1$ and $q = 4$. Analogously, when $n = 200$ we reject the null hypothesis that $q = 1$ at level $\alpha = 0.05$ in 93\%, 5\% and 100\%  of simulations when $q = 0.25$, $q = 1$ and $q = 4$. 
These rejection rates are roughly as expected given the results of the simulation study of the testing procedure given that we only perform $100$ simulations for each value of $n$ and $q$.
Figure~\ref{fig:compests_mse} shows mean squared error (MSE) for estimating $\bs \beta$ using the adaptive procedure plotted against the mean squared error for estimating $\bs \beta$ under a Laplace prior. 
We see that the adaptive procedure yields substantial improvements when the true value of $q$ is small, almost no loss when the true value of $q$ is in fact equal to $1$ and some small improvements and little loss when the true value of $q$ is large.
Smaller improvements when the true value of $q$ is large are likely due to the fact that the estimates of $q$ are more variable when $q$ is larger.
We note that incorporating testing into the adaptive procedure is important to these performance results.
Recall that both tests of $H$ based on $\psi(\hat{\bs \beta}_{ols})$ and $\psi(\hat{\bs \beta}_{\delta})$ have low power when $p$ is relatively small, i.e. when little information about the features of the distribution of $\bs \beta$ is observed.
Accordingly, incorporating testing into the estimation procedure protects us against losses in performance that could result from imprecise estimation of $q$. 

We also consider the performance of Dirichlet-Laplace (DL) priors with $a = 1/2$ and $a = 1/p$ as a comparison, which has been shown to outperform the Laplace prior in some settings \citep{Bhattacharya2015}.
Because \cite{Bhattacharya2015} assumes that $\sigma^2 = 1$, we assume that $\bs \beta/\sigma | \sigma \sim \text{DL}_a$.
We observe that the adaptive procedure posterior mean estimators outperform the posterior mean estimators under a Dirichlet-Laplace prior for both values of $a$, especially when the true value of $q$ is small.

%


\section{Relationship to Estimating Sparse $\bs \beta$}\label{sec:sparse}

\begin{figure}[h]
\centering
\includegraphics{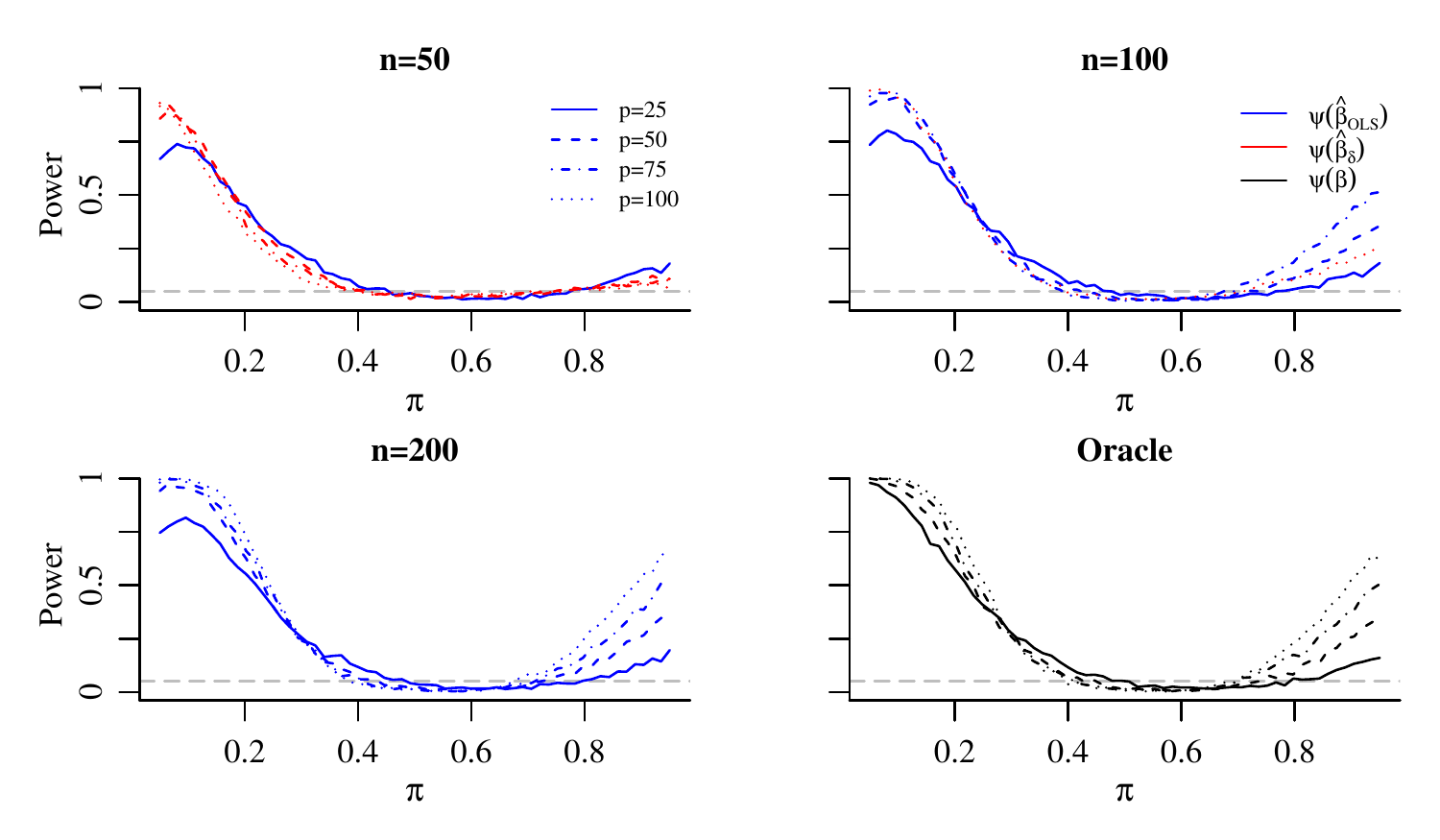}
\caption{Power and level of level-$0.05$ tests for data simulated from a linear regression model with standard normal errors and Bernoulli-normal regression coefficients with sparsity rate $1 - \pi$ and unit variance. A horizontal dashed gray line is given at $0.05$.}
\label{fig:power_misspec}
\end{figure}

The lasso penalty/Laplace prior is often used when $\bs \beta$ is believed to be sparse, i.e. many elements of $\bs \beta$ are believed to be equal to exactly zero. 
Accordingly, we repeat the testing and estimation simulation studies performed in the previous sections for Bernoulli-normal spike-and-slab distributed $\bs \beta$ where $\beta_j$ is exactly equal to zero with probability $1 - \pi$ and drawn from a $N\left(0,\tau^2/\pi\right)$ distribution otherwise.
This parametrization ensures that elements of $\bs \beta$ have variance $\tau^2$.
The kurtosis of this distribution is given by $\kappa + 3 = 3/\pi$ and when $\pi = 0.5$, the kurtosis of this distribution matches that of a Laplace distribution.
We repeat the testing simulation study in Section~\ref{sec:test} for a range of values of $\pi$ instead of $q$ and show the results in Figure~\ref{fig:power_misspec}.

As expected, our tests tend to accept $H: q = 1$ when the kurtosis of the spike-and-slab distribution is similar to the kurtosis of the Laplace prior at $\pi = 0.5$.
Importantly, our tests reject a Laplace prior when $\pi$ is small and $\bs \beta$ is very sparse.
This suggests that the adaptive procedure for estimating $\bs \beta$ might yield performance improvements even when elements of $\bs \beta$ do not have an exponential power distribution.
We repeat the estimation simulations described in Section~\ref{subsec:adaptive_simulations} for $\pi \in \left\{0.1, 0.5, 0.9\right\}$ and show the results in Figure~\ref{fig:compests_misspec_mse}. Again, histograms of estimates of $\sigma^2$, $\tau^2$ and $q$ are given in the appendix. Before discussing the results of the simulations, we note that as we might expect based on the previous simulations the sampler mixes better with larger values of $\pi$ and $n$. The smallest effective sample sizes for $n = 100$ and $n = 200$ are $70$ and $253$, respectively.

\begin{figure}
\centering
\includegraphics{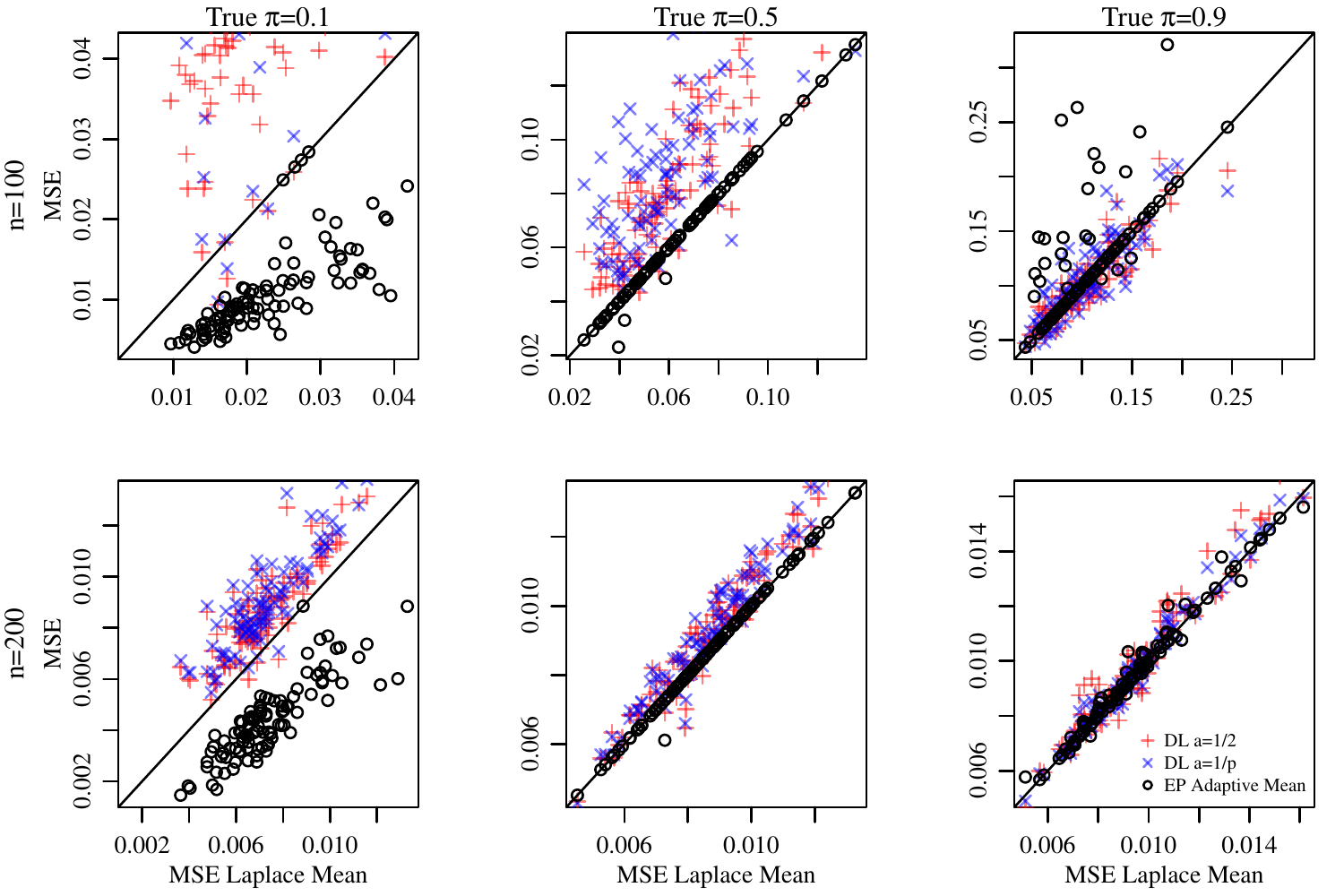}
\caption{Adaptive estimation procedure posterior mean estimator and Dirichlet-Laplace posterior mean estimator versus Laplace posterior mean estimator performance, as measured by mean squared error (MSE), for data simulated from a linear regression model with standard normal errors and Bernoulli-normal regression coefficients with sparsity rate $1 - \pi$ and unit variance.}
\label{fig:compests_misspec_mse}
\end{figure}

With spike-and-slab distributed $\bs \beta$, the adaptive procedure posterior mean estimators still outperform Laplace posterior mean estimators in the majority of simulations.   
We see substantial performance gains from the adaptive procedure posterior mean estimators relative to Laplace posterior mean estimators when $\pi = 0.1$ for both $n = 100$ and $n = 200$. 
Again, we observe some losses in performance when $\pi = 0.9$, i.e. when the kurtosis is relatively low and estimates of $q$ are more variable.
We also emphasize that incorporating the test into the adaptive procedure does play an important role in its performance. When $\pi = 0.9$ and results of a test of $H$ are ignored, the mean squared error for estimating $\bs \beta$ using $q = \hat{q}$ exceeds the mean squared error using a Laplace prior in 56\% and 67\% of simulations when $n = 100$ and $n = 200$, respectively.
When the exponential power prior is only used when a test of $H$ rejects and a Laplace prior is used otherwise, this drops to 12\% and 34\%, respectively.
We also continue to observe favorable performance of the adaptive procedure posterior mean estimators relative to Dirichlet-Laplace posterior mean estimators, especially when $\pi \leq 0.5$.

\begin{figure}
\centering
\includegraphics{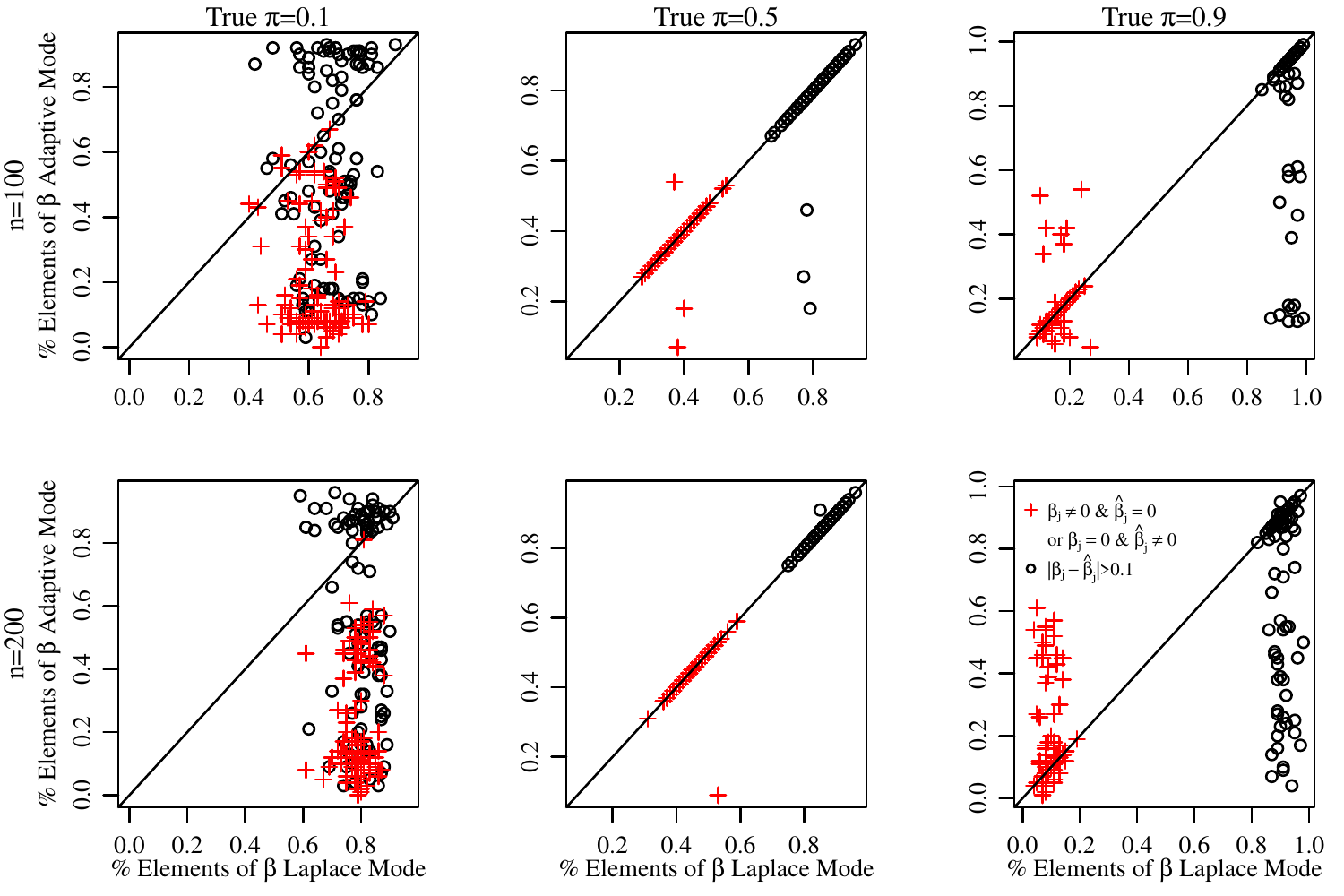}
\caption{Adaptive estimation procedure posterior mode estimator versus Laplace posterior mode estimator performance for data simulated from a linear regression model with standard normal errors and Bernoulli-normal regression coefficients with sparsity rate $1 - \pi$ and unit variance.}
\label{fig:compests_misspec_zo}
\end{figure}

Because the true values of $\bs \beta$ are sparse in these simulations, we can compare the model selection performance of the adaptive procedure posterior mode estimators versus the Laplace posterior mode estimators. Figure~\ref{fig:compests_misspec_zo} shows the proportion of zero and nonzero elements of $\bs \beta$ correctly identified, $\frac{1}{p}\sum_{j = 1}^p \mathbbm{1}_{\left\{\beta_j = 0 \text{ \& } \hat{\beta_j} \neq 0 | \beta_j \neq 0 \text{ \& } \hat{\beta}=0\right\}}$. The adaptive procedure posterior mode estimators almost always perform as well as or better than the Laplace posterior mode estimators in terms of model selection with one exception. When $n = 200$ and $\pi = 0.9$, the adaptive posterior mode estimators perform as well as or better than the Laplace posterior mode estimators in only $61\%$ of simulations (as opposed to over $\geq 97\%$ of simulations for all other values of $n$ or $\pi$). The performance of the adaptive procedure posterior mode estimators in this case likely results from the tendency to obtain estimates $\hat{q} > 1$, as the true values of $\bs \beta$ are nearly normally distributed when $\pi = 0.9$ and can be $q$ can be estimated relatively well when $n = 200$, and the unavailability of sparse posterior mode estimators under the exponential power prior when $q > 1$. As a result, we also compare the adaptive procedure and Laplace posterior mode estimators using a second metric that does not require sparsity of the adaptive procedure posterior mode estimator, $\frac{1}{p}\sum_{j = 1}^p \mathbbm{1}_{\left\{\left|\beta_j - \hat{\beta}_j\right|> 0.1 \right\}}$. Using this measure of posterior mode estimator performance, we observe that the adaptive procedure posterior mode estimator outperforms the Laplace posterior mode estimator in nearly all simulations when $\pi = 0.9$.


\section{Applications}\label{sec:app}

We apply the methods described in this paper to four datasets that have appeared previously in the penalized regression literature: the diabetes data, the Boston housing data, motif data and glucose data \citep{Efron2004, Park2008, Polson2014, Buhlmann2011, Priami2015}.
The diabetes data featured in \cite{Efron2004} contains a quantitative measure of diabetes progression for $n = 442$ patients $\bs y$ and ten covariates: age, sex, body mass index, average blood pressure and six blood serum measurements. A design matrix $\bs X$ is obtained from the ten original covariates, $10 \choose 2$ pairwise interactions and $9$ quadratic terms yielding $p = 64$. In the Boston housing data, the response vector $\bs y$ is the median house price for $n = 506$ Boston census tracts and the design matrix is made up of 13 measurements of census tract characteristics and all $13 \choose 2$ squared terms and pairwise interactions, yielding $p = 104$.
The motif data featured in \cite{Buhlmann2011} contains measurements of protein binding intensity $\bs y$ at $n = 287$ regions of the DNA sequence and $p = 195$ covariates $\bs X$ made up of measurements of motif abundance for $p$ motifs at each region.
The glucose data contains measurements of blood glucose concentration $\bs y$ for $n = 68$ subjects belonging to several families with complete data on $p = 72$ covariates, which include various metabolite measurements along with several health indicators.
We subtract an overall mean and family-specific group means off of the response and the design matrix containing the 72 covariates to be used for regression. 

For all four data sets, we centered and standardized the response $\bs y$ and the columns of the design matrix $\bs X$ by subtracting off their means and dividing by their standard deviations. We use the ridge-based test for all four data sets because either $n < p$ or the design matrix is highly collinear and $\bs X^\top\bs X$ has condition number less than $10^{-5}$. As in the simulations shown previously, we perform level-$\alpha = 0.05$ tests and approximate the corresponding quantiles $\psi_{0.025}$ and $\psi_{0.975}$ by simulating $1,000,000$ draws from the approximate distribution of the test statistic under the null. Table~\ref{tab:testres} summarizes the features of the data and the test results.

\begin{table}[h]
\centering
\begin{tabular}{ccccccc}
 \textbf{Dataset} & $n$ & $p$ & $\psi_{\delta,0.025}$ & $\psi_{\delta,0.975}$ & $\psi(\hat{\bs \beta}_{\delta})$ & $\text{Pr}\left(\psi(\bs \beta_{\delta}) \leq  \psi(\hat{\bs \beta}_{\delta})  | q = 1\right)$ \\ \hline \hline
Diabetes  & 422 & 64 &2.31 & 7.68 & 10.36 & 0.993 \\
Boston Housing &506 & 104 & 1.97 & 7.59 & 6.57 & 0.959 \\
Motif & 287 & 195 &  2.87 & 10.34 &  5.77 & 0.748 \\
Glucose & 68 & 72 & 2.31&  7.05 & 9.38 &  0.993 \\ \hline
\end{tabular}
\caption{Results of testing the appropriateness of a Laplace prior for four datasets.}
\label{tab:testres}
\end{table}

We reject the null hypothesis that a Laplace prior is appropriate for the diabetes and glucose data sets.
For these two data sets, we estimate $\sigma^2$, $\tau^2$ and $q$ and compute the posterior mode and mean estimators of $\bs \beta$ under exponential power and Laplace priors.
When computing the posterior mode estimators, we address nonconvexity when $q < 1$ by repeating the coordinate descent algorithm for $100$ randomly selected starting values and saving the estimate that gives the greatest posterior likelihood. 
Again, we caution that a unique posterior mode estimator may not exist when $\bs X$ is so highly collinear or not full rank.
We approximate posterior mean estimators using $1,000,500$ draws from each posterior distribution, discarding the first $500$ iterations as burn-in and thinning the remaining $1,000,000$ samples by a factor of $20$.

Table~\ref{tab:fitres} summarizes the variance and shape parameter estimates, mode sparsity rates and effective sample sizes for both datasets and priors.
For both data sets, estimates of the shape parameter $\hat{q}$ are less than $1$, suggesting that an even heavier tailed prior is more appropriate. 
Accordingly, the exponential power prior yields a sparser posterior mode estimator than the Laplace prior.
Mixing of the Gibbs samplers used to approximate the posterior mean estimators is better when a Laplace prior is used.

\begin{table}[h]
\centering
\begin{tabular}{cccccccc}
& \multicolumn{3}{c}{\text{Par. Ests.}}  & \multicolumn{2}{c}{\text{Mode Sparsity}} & \multicolumn{2}{c}{\text{Min. ESS}} \\
 \textbf{Dataset} & $\hat{\sigma}^2$ & $\hat{\tau}^2$ & $\hat{q}$ & $L$ & $EP$ & $L$ & $EP$   \\ \hline \hline
Diabetes  & $0.4708$ & $0.0071$ & $0.5505$  &  50.0\% & 87.5\% & 21,988 & 3,976 \\
Glucose & $0.4460$ & $0.0077$ & $0.5509$  & 80.6\% & 95.8\% &  5,545 & 782 \\ \hline
\end{tabular}
\caption{Variance and shape parameter estimates, posterior mode estimator sparsity rates and minimum effective sample sizes of posterior samples under Laplace (L) and exponential power (EP) priors.}
\label{tab:fitres}
\end{table}

\begin{figure}[h]
\centering
\includegraphics{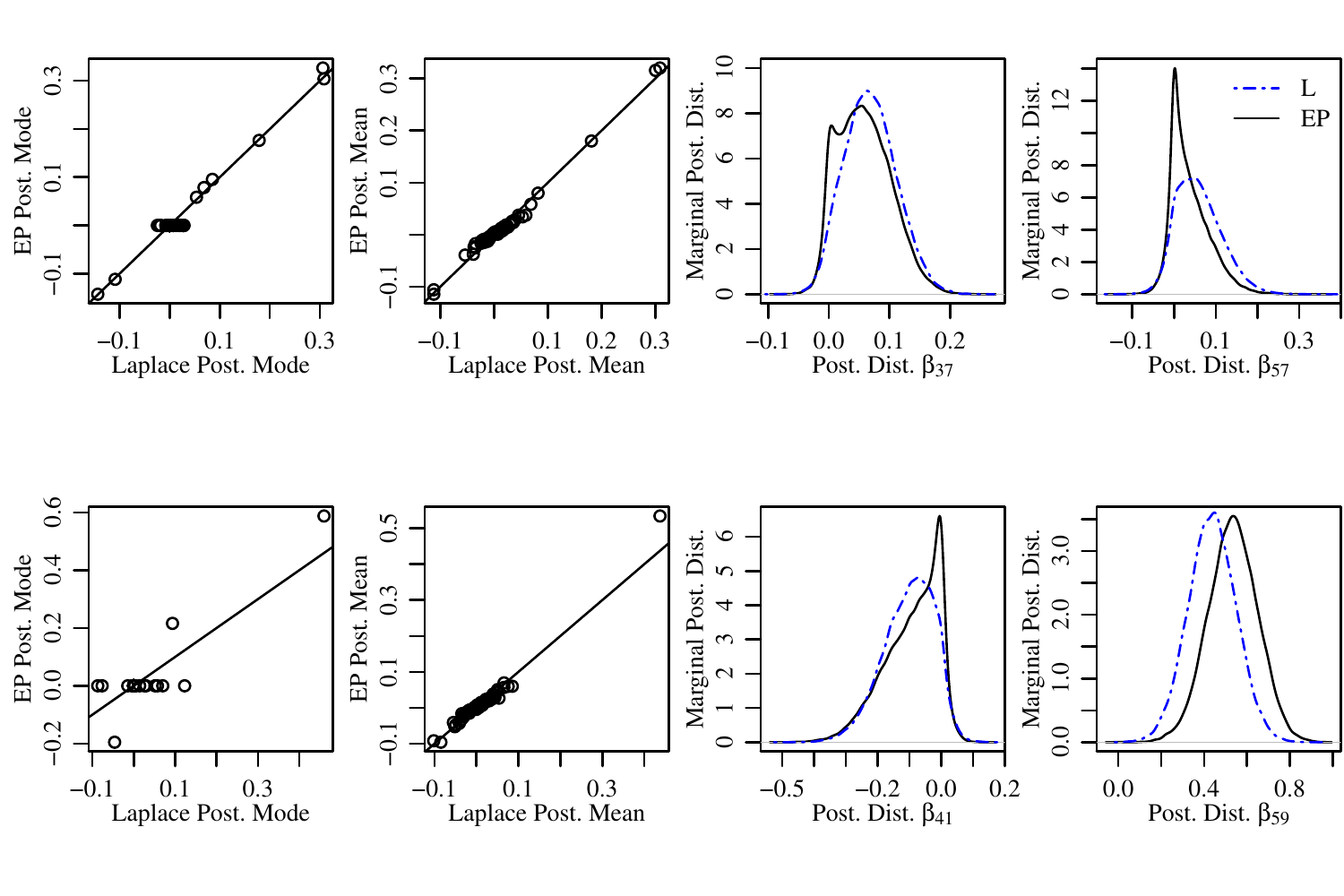} 
\caption{Posterior modes, means and selected marginal distributions under exponential power (EP) priors and Laplace (L) priors of $\bs \beta$ for diabetes and glucose datasets.}
\label{fig:fitvis}
\end{figure}

Figure~\ref{fig:fitvis} compares posterior mode and mean estimators and selected marginal distributions under Laplace and exponential power priors for $\bs \beta$.
Examining the posterior mode estimators, we observe not only higher sparsity rates but also less shrinkage of nonzero values when the exponential power prior is used.
We observe similar but less stark differences when comparing posterior mean estimators across both priors. 
We also compare the marginal posterior distributions for several elements of $\bs \beta$, chosen to demonstrate how using an exponential power prior affects inference for these datasets.
In the right four panels of FIgure~\ref{fig:fitvis}, we see that using the exponential power prior can cause the mode of the marginal posterior distribution to change locations or can introduce bimodality of the marginal posterior distribution.
Overall, we gain more interpretable estimates of $\bs \beta$ with fewer large entries by using a more appropriate prior.

\section{Discussion}\label{sec:disc}

In this paper, we introduce a simple procedure for testing the null hypothesis that a Laplace prior is appropriate by assessing whether or not the kurtosis of the distribution of unknown regression coefficients matches that of a Laplace distribution.
We also introduce two-step adaptive estimation procedure for $\bs \beta$ that uses an exponential power prior for $\bs \beta$ if a Laplace prior is rejected.
We show that our testing and estimation procedures perform well for the kinds of values of $n$ and $p$ we might encounter in practice both when elements of $\bs \beta$ have an exponential power distribution and when they are sparse with a Bernoulli-normal spike-and-slab distribution.
We have demonstrated that the appropriateness of a Laplace prior for estimating Bernoulli-normal spike-and-slab $\bs \beta$ depends on the sparsity rate and that estimates based on a Laplace prior can be suboptimal when we expect that $\bs \beta$ follow a spike-and-slab distribution with a high sparsity rate.
As dependence of kurtosis on the sparsity rate is not limited to the Bernoulli-normal spike-and-slab distribution but rather extends to any spike-and-slab distribution where the slab is a mean zero distribution with finite fourth moments, we expect that the performance improvements we observe might persist for more general sparsely distributed $\bs \beta$.
This compliments the existing statistical literature on the suboptimality of the Laplace prior \citep{Griffin2010, Carvalho2010, Polson2014, Bhattacharya2015}.

This work has several natural extensions.
Because the derivation of the approximate level-$\alpha$ test follows from the existence of a consistent estimator of $\bs \beta$ or a known linear function of $\bs \beta$ the methods described in this paper can be extended to include linear models with elliptically contoured errors and generalized linear models.
The methods described in this paper could also be extended to include the construction of a confidence interval either for the kurtosis of the distribution of the elements of $\bs \beta$ or the exponential power shape parameter $q$. 
Additionally, the favorable performance of the adaptive estimation procedure suggests that the construction of a likelihood ratio or Wald test of the appropriateness of the Laplace prior and likelihood-based estimation of $\sigma^2$, $\tau^2$ and $q$ may be worth revisiting.
 We expect that likelihood-based tests and estimators might be more powerful and efficient than the moment-based tests and estimators we introduce, if they could be obtained in practice.
Future work might consider how advances in empirical Bayes estimation of hyperparameters, such as  \cite{Doss2010}, might be used to overcome the computational challenges that make maximum likelihood estimation of $\sigma^2$, $\tau^2$ and $q$ and the development of  likelihood-based tests prohibitively computationally demanding.
Furthermore, the methods can be generalized to test the null hypothesis that elements of $\bs \beta$ have an exponential power distribution with $q = \tilde{q} >0$ or to test a null hypothesis that elements of $\bs \beta$ have a different symmetric, mean zero distribution as long as this different distribution can be characterized by its kurtosis and is easy to simulate from, e.g. the normal-gamma distribution given by \cite{Griffin2010} or the Dirichlet-Laplace distribution given by \cite{Bhattacharya2015}.
Last, throughout this paper we have conflated heavy tails (high kurtosis) with ``peakedness'' of the density of $\bs \beta$. 
However, it is not generally true that ``peakedness'' must increase with tail weight  \citep{Westfall2014}. 
This is important because sparsity of the posterior mode estimator specifically arises from the ``peakedness'' of the prior on $\bs \beta$.
Accordingly, three parameter distributions like the generalized $t$-distribution given by \cite{Choy2008} that allow kurtosis and ``peakedness'' to vary separately may be useful alternative priors for $\bs \beta$.

\bigskip
\begin{center}
{\large\bf SUPPLEMENTARY MATERIAL}
\end{center}

Supplementary material available online includes proofs of all the propositions, coordinate descent and Gibbs sampler details and additional numerical results. R code for implementing the methods described in this paper is also available, including packages for implementing coordinate descent and Gibbs sampling for mode estimation and posterior simulation under an exponential power prior. Data is also included as supplementary material.

\bibliographystyle{chicago}
\bibliography{EBRegression}

\end{document}